\documentclass[
twocolumn,
aps,prd,
nofootinbib,
superscriptaddress,
tightenlines,
amsmath,
amssymb
]{revtex4}
\usepackage{bm}
\usepackage{color,graphicx}
\usepackage{amsmath}
\usepackage{amssymb}

\newcommand{\be}{\begin{equation}}
\newcommand{\ee}{\end{equation}}
\newcommand{\bea}{\begin{eqnarray}}
\newcommand{\eea}{\end{eqnarray}}
\begin{document}

\title{Partonic Interpretation of Generalized Parton Distributions}

\author{Gary R.~Goldstein} 
\email{gary.goldstein@tufts.edu}
\affiliation{Department of Physics and Astronomy, Tufts University, Medford, MA 02155 USA.}
\author{Simonetta Liuti} 
\email{sl4y@virginia.edu}
\affiliation{Department of Physics, University of Virginia, Charlottesville, VA 22901, USA.}
\begin{abstract}
Generalized Parton Distributions describe, within QCD factorization, the non perturbative 
component in the amplitudes for deeply virtual exclusive processes.  
However, in order for a partonic 
interpretation to hold, semi-disconnected diagrams should not contribute.
We show that this condition is not satisfied for non-forward kinematics at leading order, and that gluon mediated 
re-interactions are essential for a consistent description in terms of parton degrees of freedom. 
\end{abstract}


\maketitle

\baselineskip 3.0ex
%%%%%%%%%%%%%%%%%%
Generalized Parton Distributions (GPDs) are defined as extensions of the parton distributions from Deep Inelastic Scattering (DIS) to a more complex phase space domain where off-diagonal matrix elements can be related to the partons displacements in transverse space. 
As a consequence of the factorization theorem of QCD \cite{Ji}, GPDs enter the matrix elements for the Deeply Virtual Compton Scattering (DVCS) amplitudes.
% depicted in Figure \ref{fig0}.  
 Because the momenta of the outgoing and incoming quark and proton are different, two distinct kinematical regions can be defined. By denoting $X=k^+/P^+$ the light cone momentum fraction of the struck quark relative to the initial proton momentum, $P^+$, and by $X-\zeta = k^{\prime +}/P^+$ the corresponding momentum fraction of the returning quark ($\zeta = \Delta^+/P^+$ represents the t-channel momentum transfer fraction), two distinct regions appear.  In the $X > \zeta$ region both the struck and returning quark carry positive momentum fractions of the initial proton momentum. This is called the Dokshitzer-Gribov-Lipatov-Altarelli-Parisi (DGLAP) region because of the way parton evolution is expected to proceed. 

The $X < \zeta$ region has been interpreted as describing a quark-antiquark pair emerging from the proton, more similar to the generalization of a distribution amplitude. Evolution proceeds through the Efremov-Radyushkin-Brodsky-Lepage (ERBL) mechanism.  
 
This way of interpreting the ERBL region was proposed at the inception of DVCS studies. In
%However, upon further in depth studies of the analytic properties of the DVCS amplitudes 
\cite{GolLiu_disp} 
we found a reason of concern in noticing that  it
imposes several seemingly artificial constraints  for partonic based descriptions and model building. 
In particular, symmetry requirements under $X \rightarrow -X$ follow for charge conjugation, $C$, even or odd combinations of parton distributions. These violate the standard Kuti-Weisskopf separation of valence (flavor non singlet) and sea (flavor singlet) quarks, and they may not be naturally satisfied in a large class of models including all spectator models \cite{AHLT,BroEst,Metz}.
An even more compelling issue examined here is whether the ERBL region relates at all to the proton's partonic substructure. 
In order to prove this, one has to ascertain that the quark anti-quark pair emerges directly from the proton,  
rather than being a vacuum hadronic fluctuation.     
%Work in $A^+=0$ gauge, LC definitions.
%%%%%%% Figure 0
%\begin{figure}
%  \includegraphics[width=12cm]{Jaffe_parton_0.pdf}
%  \caption{DVCS amplitude}
%  \label{fig0}
%\end{figure}
%%%%%%%
%
%What defines the partonic content of the proton? 
Three conditions are essential  to characterize a partonic description \cite{Jaffe}:

\begin{enumerate}
\item  the support in $X$ is defined by the region  $\mid X \mid \leq 1$;
\item analytic properties of the partonic amplitude have to correspond to the emission and absorption of quarks/antiquarks 
via well defined on-mass shell intermediate hadronic states;
\item the quark-proton vertices have to be connected.
\end{enumerate}

A careful derivation of the parton model from the connected matrix elements for the non-local quark and gluon fields operators that enter inclusive hard processes was given in \cite{Jaffe} (a formal extension to the off-forward case was given in \cite{Die_Gou}). There it was pointed out that a ``simple" physical picture does not emerge uniquely and naturally from the structure of the correlator, but that analytic properties need to be taken into consideration. 
%While this is straightforward for twist-2 operators, when more partons are involved as {\it e.g.} with twist-3, these conditions become essential.   

We now extend the arguments of \cite{Jaffe} to the ERBL region that similarly presents a more complicated partonic structure.  
%We start from the hadronic tensor
%\begin{equation}
%T_{\mu \nu} = i \int d^4 z  e^{i q z} \langle P^\prime  \mid T(J_\mu(z) J_\nu(0)) \mid P \rangle 
%\end{equation}  
%
A factorized  form was derived in Ref.\cite{Ji} (for a review see \cite{Diehl_rev}) 
for the DVCS amplitude as
\begin{eqnarray}
\mathcal{F} & = & \int\limits_{-\zeta+1}^{1} dX \left(\frac{1}{X-\zeta + i \epsilon} - \frac{1}{X+i \epsilon} \right)  \nonumber \\
& \times&  \int d z^- e ^{i \, q^+ z^-} \langle P^\prime \mid \bar{\psi} (z^-) \gamma^+ \psi(0) \mid P \rangle. 
\label{correlator}
\end{eqnarray}
The matrix element in the equation corresponds to GPDs defined {\it e.g.} in the unpolarized case as
\begin{eqnarray}
\label{correlator}
\int d z^- e ^{i \, q^+ z-} \langle P^\prime \mid \bar{\psi} (z^-) \gamma^+ \psi(0) \mid P \rangle = & & 
\nonumber \\
\overline{U}(P^\prime) \left[ H(X,\zeta,t) \, \gamma^+ + E(X,\zeta,t) \, \frac{-i \sigma^{+,\lambda}}{2M} \Delta_\lambda \right] U(P) & & 
\end{eqnarray}
%For simplicity we will focus on the proton non spin flip GPD, $H$.

To clarify the identification of partonic or non-partonic interpretations of the GPDs we can explicitly expand the quark field operators of  Eq.(\ref{correlator}) in terms of light front free field variables (implicitly employing the operator product expansion as in Ref.~\cite{Jaffe}).
We use the decomposition into creation and annihilation operators
%
%%% 
%\footnote{We will not use $\xi\ge 0$, which takes advantage of the symmetry or antisymmetry under $\xi\rightarrow -\xi$, a property that may not be satisfied in particular spectator models.\cite{Golec}.} 
%%%
%%%
focusing for simplicity on $H$, as \cite{Diehl_rev}
%%%
\begin{eqnarray}   
\label{H_GPD}   
& & H(X, \zeta,t)  =  \frac{1}{2 \bar{P}^+}  \sum_\lambda \int  \frac{d^2 k_T}{2 
\sqrt{\mid X^2 - \zeta^2  \mid} (2 \pi)^3}     
\nonumber \\    
& &  \left[  \langle P^\prime | b_\lambda^\dagger \, (X-\zeta, {\bf k}^\prime_T ) \: b_\lambda    
\, (X, {\bf k}_T) | P \rangle \:  
\right .    
\nonumber 
\\  
& & 
+ \;  \langle P^\prime \mid b_\lambda^\dagger \, (X-\zeta 
, -{\bf   k}_T ^\prime)  
\: d_{- \lambda}^\dagger \, (X,    
{\bf k}_T) | P \rangle 
 \nonumber  
\\   
& & + \; \langle P^\prime | d_\lambda \, (X-\zeta, - {\bf  k}^\prime_T )  
\: b_{ -\lambda} \, (X,    
{\bf k}_T) | P \rangle \: 
\nonumber
\\
& & 
+ \; \left . \langle P^\prime | d_\lambda \,, (X-\zeta,    
{\bf k}^\prime_T) \: d_\lambda^\dagger \,  
(X, {\bf k}_T) | P \rangle \: 
%\theta (x \le -\xi)  
\right ].     
\end{eqnarray}  
%The first  (fourth) line represents creation and annihilation of quarks (antiquarks) on both sides of the diagram, the second and third line represent the creation of a quark-antiquark pair.

\begin{figure}
  \includegraphics[height=7cm]{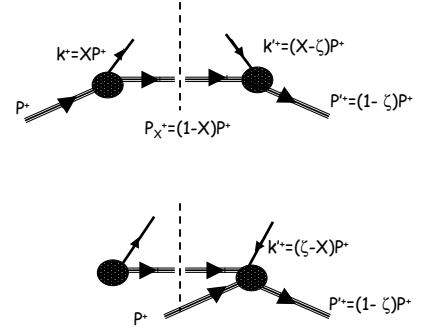}
  \caption{Upper Panel: Parton Model; Lower Panel:  Semi-disconnected contribution. 
 The analytic structure of these diagrams was discussed for DIS in Ref.\cite{Jaffe}.}
  \label{fig1}
\end{figure}

The relation to possible on-shell intermediate states can be explored by inserting a complete set between the quark field operators acting between the incoming or outgoing proton states. Whether or not the corresponding diagrams contribute depends on the values of $X$ and $\zeta$ and the momenta. It was pointed out by Jaffe~\cite{Jaffe} that even in the forward case,  when $\zeta=0$ and $X<0$ there are semi-disconnected contributions to the ``unitarity diagrams'', shown in Fig.~\ref{fig1}, that do not have a partonic interpretation. However, there are two equivalent forms of the product of the two non-local interacting quark field operators, using the anti-commutation of the operators on the null-plane
% (a ``remarkable'' result from which it follows that the T-product is ``illusory''~\cite{Jaffe}) . 
As a result, from the equivalence
of these two forms on can deduce an equivalence between the non-partonic/semi-disconnected  diagrams with $X<0$ for {\em quarks} and, through the alternative ordering of the fields, a partonic distribution for {\em anti-quarks} with $X>0$ represented by the usual connected configuration.
%spectator intermediate states that get summed over. 
The important question here is whether or not this kind of equivalence can be established in the off-forward case in order to allow a partonic interpretation of GPDs.

%%%%%%% Figure 1
%%%%%%%

To consider these questions, insert  intermediate states in Eq.(\ref{correlator}) using completeness, 
%$\Sigma_n | n \rangle\langle n | = 1$, 
and associate each vertex with plus momentum conservation. 
%Then each term 
%carries momentum defined by 
%\[ \delta(P^+ - XP^+ - P_n^+) \]  or equivalently,  \[ \delta((1-\zeta)P^+ - (X-\zeta)P^+ - P_n^+). \]
In the forward limiting case one obtains
\begin{equation} 
H(X,0,0) = \sum_n \delta(P^+ - XP^+ - P_n^+) \mid \langle  n \mid \psi \mid P \rangle \mid^2
\label{H_X00}
\end{equation}
so for quarks with $X>0$ this corresponds to the usual parton picture in Fig.~\ref{fig1}a. For $X<0$ the delta function requires that the intermediate state momentum exceeds the proton $p^+$, that is $p_n^+=P^+ + |X|p^+$. That condition can be satisfied by the creation of a pair from the vacuum as in Fig.~\ref{fig1}b, as well as  through annihilation of a pair into the vacuum, as shown in detail by Jaffe~\cite{Jaffe}. 

%For the GPD in Eq.(\ref{H_GPD}), 
%$H$ becomes an off-diagonal matrix element for the exclusive amplitude,
Similarly, in the off-forward case
\begin{equation}
H(X,\zeta,t) = \sum_n \delta(P^+ - XP^+ - P_n^+) \langle P^\prime \mid {\bar \psi} \mid n \rangle \langle n \mid \psi \mid P \rangle 
\end{equation}
%for anti-quarks, with $X<0$. 
%$\psi^+$ is the (good) component (explain? redundant?) of the function.  
By expanding $\psi$ and 
concentrating on the second term in Eq.(\ref{H_GPD}) to illustrate the procedure, we have
%\begin{subequations}
\begin{eqnarray}
%& & \; \langle P^\prime | b^\dagger_\lambda \, ((X - \zeta)  
%\bar{P}^+, -\mbox{\boldmath   $k$}^\prime_T ) | n \rangle \langle n |
%\: b_{\lambda} \, (X \bar{P}^+,    
%\mbox{\boldmath $k$}_T) | P \rangle \: \delta(P^+ - xP^+ -P_n^+)
%\label{first} \\
%%%
& &  \langle P^\prime | d_\lambda \, (X - \zeta, -\mbox{\boldmath   $k$}^\prime_T) \mid n \rangle \langle n \mid
\: b_{- \lambda} \, (X,    
\mbox{\boldmath $k$}_T) | P \rangle   \nonumber  \\
& & \times \delta(P^+ - XP^+ -P_n^+),
\label{second}
\end{eqnarray}
%\end{subequations}
that
annihilates a quark with $X P^+$ 
%or $(x+\xi)P^+$ 
and creates an antiquark with $-(X-\zeta)P^+$. 
%or $-(x-\xi)P^+$. 
%
For the DGLAP region this cannot conserve plus momentum, 
%(say that $X>0$ differently from the fourth term). 
%
but for the ERBL region, $-(X-\zeta)P^+$ corresponds to an antiquark. 
Notice that the delta function has $P_n^+$ greater than $P^+$.

%a situation that can be satisfied if the antiquark and the intermediate states, with quark quantum numbers, are created from the vacuum, as in the figure.

More heuristically, each of the terms in Eq.(\ref{H_GPD}) can, on one side describe the processes
\[ p \rightarrow q (qq) \; \; \; {\rm and} \; \; \;   q  (qq) \rightarrow p \] 
respectively, at each vertex.  The intermediate state $(qq)$ has the quantum numbers of a diquark with momentum $P_n$. 
The equation has a clear partonic interpretation.
%$X<0$ corresponds either to the anti-quark case (write better). 
On the other side, for Eq.(\ref{second}), consistently with momentum conservation,  one has two different types of intermediate states:
{\it i)} $p \rightarrow  q(qq q\bar{q})  $ on the LHS, the $\bar{q}$ being re-emitted on the RHS;
{\it ii)}  
$ p \rightarrow q p (\bar{q})  $ on the LHS, the $\bar{q}$ being re-emitted through the semi-disconnected vertex of Fig.{\ref{fig1}b. 
In case {\it i)} the intermediate state has diquark quantum numbers, a partonic interpretation seems possible but with a catch that we explain in what follows. 
In case {\it ii)} the intermediate state is a $\bar{q}$.
The semi-disconnected graphs do not correspond to a partonic description of the proton. 

One can show how each of the four terms in Eq.~(\ref{H_GPD}) corresponds to connected or semi-disconnected graphs. Writing the matrix element for the second term, when $X<\zeta$
\bea
 & &\langle P^\prime \mid b_\lambda^\dagger \, (X- \zeta, -{\bf   k}_T^\prime)  | n \rangle \langle n|
\: d_{- \lambda}^\dagger \, (X,    
{\bf k}_T) | P \rangle =
\nonumber \\
& & \langle P^\prime \mid b_\lambda^\dagger \, (X- \zeta, -{\bf   k}_T ^\prime )  |P, n \rangle \langle n|
\: d_{- \lambda}^\dagger \, (X,    
{\bf k}_T)| 0\rangle  
%| P \rangle
\eea
where the $b^\dagger$ has plus momentum $-|X-\zeta|P^+$. This corresponds to the Fig.~\ref{fig1}b, a ``semi-disconnected'' graph. In the DIS case, the identification of the initial and final matrix elements in Eq.(\ref{H_X00}) allows the replacement of the semi-disconnected quark target diagrams for $X<0$ with the connected antiquark-target diagram for $X>0$ (with opposite sign). On the other hand for the GPD, because of the asymmetry between the initial quark-target state and the final state, the semi-disconnected diagrams for $X-\zeta<0$ are equivalent to {\it semi-disconnected diagrams} for antiquark-target states, {\it i.e.} the ERBL region for quark-target amplitudes is the ERBL region for anti-quark target amplitudes.

Hence, there is a catch that casts a doubt on the possibility of giving a partonic interpretation of the ERBL region. By examining the analytic structure of GPDs, we know that in this region it is 
either one of the struck quarks/anti-quarks that is put on mass shell  \cite{BroEst}, and not the state with  diquark  quantum numbers.
The only way to have a $\bar{q}$ or equivalently a $q$, as an intermediate is by considering 
the diagram as in Fig.\ref{fig1}b \cite{Jaffe}, a semi-disconnected, non-partonic contribution. This development contradicts the partonic interpretation of the results for the ERBL region 
%based on 
%a simple covariant scalar quark and diquark-spectator model ({\it e.g.} Ref.~\cite{BroEst}), 
where the Cauchy integration over the quark momentum puts the quark on-shell in the ERBL region, while having an off-shell diquark. 
In other words, the interpretations of the ERBL region, as would be obtained in \cite{Golec,Die_Gou} by inserting different creation and annihilation operators does not address the issue of partonic interpretation. These would  require diquark-type states as the intermediate states even in the ERBL region. 
%Everything seems consistent  from the momentum balance point of view if one takes the diquark as an intermediate state in the sense that one gets $p_n^+<p+$ on the LHS and $p_n^+ > p^+$ on the RHS and this is fine.

In summary, only semi-disconnected graphs contribute to the ERBL region at leading order (Fig.\ref{fig1}). These, in turn, do not correspond to partonic distributions.  We have a choice. We can take their contribution as $H^{(2)}_{X<\zeta}=0$.
% where the index $2$ refers to the fact that only two partons (or a parton and anti-parton) are involved. 
%This proves that as soon as $\zeta \neq 0$ there is trouble giving a partonic description. 
%%%% NEW 
%%%% Connection with factorization
%Notice that our result is consistent with QCD  factorization.
% because the latter does does not specify the partonic content of the non-perturbative piece.
Alternatively we can conclude that there are non-partonic contributions to the GPDs and measurements of the ERBL region are not revealing the partonic content of the nucleons, or even the distribution of quark-antiquark meson states in the nucleon.  

The impasse in trying to give a partonic interpretation of the ERBL region could 
be overcome by considering multiparton configurations {\it i.e.} extending the definition (\ref{correlator})  to more than two parton fields. Following Ref.\cite{Jaffe} one can write 
diagrams of the type represented in  Fig.\ref{fig2}. The corresponding analytic structure is given by
%%%%%%%
%%%%%%% Figure 2
\begin{figure}
  \includegraphics[height=6.5cm]{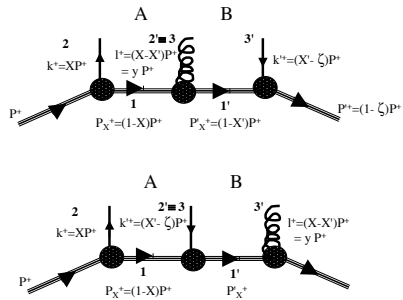}
  \caption{Multiparton contributions to DVCS. Upper panel: a connected  one gluon contribution; Lower panel: a connected non planar contribution.}
  \label{fig2}
\end{figure}
%%%%%%%
%%%%%%%%
\begin{eqnarray}
& & H^{(3)}(X,X^\prime,\zeta,t)  =  \sum_{n,m} \delta(P^+ - XP^+ - P_n^+) \nonumber \\
& & \times \delta(P^+ - X^\prime P^+ - P_m^+)  \nonumber  \\
& & \times \langle P^\prime \mid \bar{\phi} \mid n \rangle \langle n \mid \phi \mid m \rangle  \langle m \mid \phi \mid P \rangle 
\label{multiparton}
\end{eqnarray}
These configurations allow us to describe the ERBL region in terms of connected diagrams as we explain below (a more detailed discussion and model evaluations are in progress \cite{GolLiu_prog}).   
We distinguish two loops, $A$ (left) and $B$ (right), and locate the poles on the complex  $k^-$ and $k^{\prime \, -} $ planes, with longitudinal variables $X$ and $X^\prime - \zeta$, respectively. 
%The positions of the poles relative to the imaginary axis is determined by using Light Cone (LC)  coordinates and by considering quantities of the type
%\[ k^- = \frac{\Lambda^2 + k_\perp^2 -i \epsilon}{XP^+} \]  
%where $\Lambda^2$ is a $k_\perp$ independent mass term. Note that what determines the Cauchy integration, or the sign of the imaginary component in the equation, is the sign of the ``$+$" component of the parton's momentum. 
We first evaluate the position of the poles in loop $A$. This is similar to the two-particle case examined before.
We then consider only the contribution from positive $+$ momentum for the intermediate particle, since only this will give a connected diagram. The calculation of the poles for loop $B$ is described in Fig.\ref{fig3}.  
%Notice that now the Cauchy integration momentum variable is $k^{\prime \, -}$.
%the final quark's "forward" momentum.
%%%
In the figure we represent the positions of the poles for loop $B$ in both the planar (upper panel) and non planar (lower panel) cases. The gluon's momentum fraction is $y=X-X^\prime$. In loop $B$ there is much more flexibility in the position of the poles since now both the intermediate gluon and the returning quark plus momentum components can change sign.  One can see that similarly to the simpler case of Fig. \ref{fig1}, only the contributions where parton $1'$ is on shell correspond to a connected diagram and therefore to a partonic configuration. At variance with the simpler two-parton configuration examined before, this is now possible in the ERBL region ($X-\zeta<0$) as one can see by inspecting the non-planar configuration in Fig.\ref{fig3} (lower panel, left). The kinematics for that placement of poles is as follows: $X-X'<0$, $X'>\zeta$, for which the combination $X-\zeta = (X-X') + (X'-\zeta)$ will be $<0$ when the gluon carries away a larger momentum fraction than the returning quark brings back. This is clearly a connected contribution to the ERBL region. %of the GPDs. %For the other situation, where $X-\zeta>0$ the poles are all on the same side of the $k^{\prime \, -}$ plane, leading to a vanishing contribution.
 
In summary, multiparton configurations involving one exchanged gluon allow for more flexibility in the division of hard momenta - an extra loop integration is performed. The resulting calculation of the corresponding Cauchy integrals does not vanish when the quark-gluon combined momenta ($X-\zeta<0$) are in the ERBL region, {\it i.e.} when the combination is propagating like a dressed antiquark. (Such a combination would have a threshold cut in the variable $k^{\prime \, 2}$.) This provides the first, lowest order non-zero connected contribution to GPDs in the ERBL region.  
Whether this configuration lends itself to a simple partonic interpretation is related to the issue of whether parton distributions are probability distributions \cite{Bro_sannino} (see also Ref.~\cite{BJY}), and of whether such multiparton contributions can be considered to be Final State Interactions (FSI) 
of the colored separated states, and thereby are not suppressed by powers of the hard scale \cite{BHS}.
%, as was shown in the TMD models beginning with Ref.~\cite{BHS}. We will address this and other related questions in a forthcoming publication~\cite{GolLiu_prog}.
   
%%%% fino a qui
%In summary, one can write either the fully connected graph (with $x>0$ for quarks and $x<0$ for the antiquark case), or the sum of the three semi-disconnected graphs (with $x<0$ for quarks and $x>0$ for the antiquark case). (We  have to write a proof of this in formulae?). 
%The parton model follows from the fully connected graphs. So for $x<0$ the three semi-disconnected graphs in Jaffe can be replaced with the anti-quark parton model graph. Viceversa for $x>0$. The parton model is always valid. 

%%%%%%% Figure 3
\begin{figure}
  \includegraphics[height=0.25\textheight]{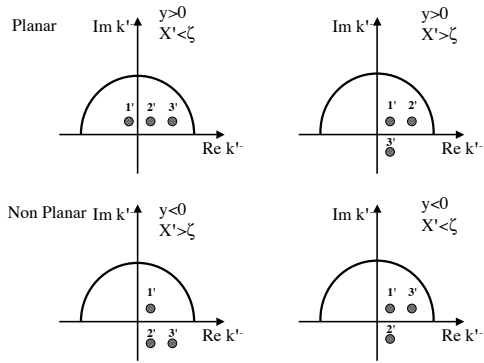}
  \caption{Representation of poles in the integration variable, $k^-$, complex plane for the multiparton diagrams of Fig.\protect\ref{fig2}.}
  \label{fig3}
\end{figure}
%%%%%%%
%%% LF
We finally notice that the complications arising within partonic descriptions in the ERBL region were also addressed using Light Front  Quark Models (LFQM) \cite{CRJi_01,CRJi_06,BroHwa,BroDieHwa,Miller,Salme}. 
In LFQM the contribution to the ERBL region is identified with  ``non-valence"  type diagrams, which are also related to multiparton configurations, or higher Fock components in the hadronic wave function, 
%However, a fully consistent connection with QCD is harder to define in that 
and semi-disconnected diagrams do not appear. These are instead reinterpreted as an analytic 
continuation of the Bethe-Salpeter (BS) wave function, thus suggesting a simpler vacuum structure. 
However, a fully consistent connection with QCD would require that both an explicit treatment of hard rescattering contributions in the amplitude \cite{Braun}, and, most importantly, of the analyticity properties are addressed.   
In this paper we showed that in order to establish the correct support, crossing symmetry, and analyticity properties that are necessary to establish dispersion relations, hence a partonic interpretation of GPDs, 
one needs a description beyond the identification of the proton off-forward structure functions with partonic wave functions. This can be accomplished by bringing into play FSI.   

We thank R.L. Jaffe for helpful discussions. 

%%%%%%%%%% OBSERVATIONS
%\vspace{0.5cm}%
%\noindent {\bf Observations (to be erased in publishable version})
%\begin{enumerate}
%\item We have not proven that the $T$ product of currents can be replaced by fields in any order, as done in \cite{Jaffe} in forward case. We assume it is true. Then all the analytic properties follow (S.L.).  
%\end{enumerate}

%%%%%%%%%%%%%%%%%%

%%%%%%%%%%%%%%%%%%
\end{document}